\documentclass[twocolumn,showpacs,preprintnumbers,amsmath,amssymb]{revtex4}

\usepackage{graphicx}
\usepackage{dcolumn}
\usepackage{bm}

\begin{document}


\title{Whistleron Gas in Magnetized Plasmas}

\author{Salvatore De Martino}
 \email{demartino@sa.infn.it}

\author{Mariarosaria Falanga}%
 \email{rosfal@sa.infn.it}

\author{Stephan I. Tzenov}
 \email{tzenov@sa.infn.it}

\affiliation{%
Dipartimento di Fisica "E.R. Caianiello", Universit\'a degli Studi
di Salerno and INFN Sezione di Napoli
\\Gruppo Collegato di Salerno, Via S. Allende, I-84081 Baronissi (SA), Italy}%

\date{\today}

\begin{abstract}
We have studied the nonlinear dynamics of whistler waves in
magnetized plasmas. Since plasmas and beam-plasma systems
considered here are assumed to be weakly collisional, the point of
reference for the analysis performed in the present paper is the
system of hydrodynamic and field equations. We have applied the
renormalization group method to obtain dynamical equations for the
slowly varying amplitudes of whistler waves. Further, it has been
shown that the amplitudes of eigenmodes satisfy an infinite system
of coupled nonlinear Schr\"{o}dinger equations. In this sense, the
whistler eigenmodes form a sort of a gas of interacting
quasiparticles, while the slowly varying amplitudes can be
considered as dynamical variables heralding the relevant
information about the system. An important feature of our
description is that whistler waves do not perturb the initial
uniform density of plasma electrons. The plasma response to the
induced whistler waves consists in velocity redistribution which
follows exactly the behaviour of the whistlers. In addition,
selection rules governing the nonlinear mode coupling have been
derived, which represent another interesting peculiarity of our
description.
\end{abstract}


\maketitle

\vspace{10. mm} KEY WORDS:   Magnetized Plasma, Renormalization
Group, Whistler Waves, Solitary Waves.

\renewcommand{\theequation}{\thesection.\arabic{equation}}

\setcounter{equation}{0}

\section{\label{Intro}Introduction}

Over four decades passed since it was first shown that plasmas and
beam-plasma systems immersed in an external magnetic field can
support travelling electromagnetic waves with specific features.
These waves propagate parallel to the applied magnetic field being
circularly polarized in a plane transverse to the direction of
propagation. It has become conventional in the physics of
magnetized plasmas to call such structures waves in the whistler
mode.

Although the linear stability properties of the electromagnetic
waves in the whistler mode are relatively well studied
\cite{Weibel,Neufeld,Bell,Sudan}, there is a serious gap in the
understanding of their nonlinear behaviour. Chen et al.
\cite{Wurtele} have shown that electromagnetic whistler waves can
be considered as complementary to the nonlinear travelling
electrostatic waves, known as the Bernstein-Greene-Kruskal (BGK)
modes \cite{BGK}. While the BGK modes are longitudinal, the
whistler modes are transverse, in other words, the components of
the electric and magnetic field of the whistler wave parallel to
the external magnetic field are both zero. The study of the
nonlinear behaviour of whistler waves has been initiated by
Taniuti and Washimi \cite{Taniuti}, who obtained a nonlinear
Schr\"{o}dinger equation for the slowly varying amplitude (see
also Reference \cite{Shukla}).

The present paper is aimed at filling the gap in the understanding
of the nonlinear evolution of whistler waves. The method adopted
here is the renormalization group (RG) method \cite{Oono,Tzenov}.
The basic feature of this approach is that it provides a
convenient and straightforward tool to obtain an adequate
description of the physically essential properties of
self-organization and formation of patterns in complex systems.
Coherent structures which result from the nonlinear interaction
between plane waves evolve on time and/or spatial scales
comparatively large compared to those the fast oscillations occur.
The RG method can be considered as a powerful systematic procedure
to separate the relatively slow dynamics from the fast one, which
is of no considerable physical relevance.  In a context similar to
that of the present paper, it has been successfully applied by one
of the authors \cite{Tzenov,Tzenov1} to study collective effects
in intense charged-particle beams.

The paper is organized as follows. In the next section, we state
the basic equations which will be the subject of the
renormalization group reduction in section III. Starting from a
single equation [see equation (\ref{Waveeqallord})] for the
electromagnetic vector potential, we obtain a formal perturbation
expansion of its solution to second order. As expected, it
contains secular terms proportional to powers of the time variable
which is the only renormalization parameter adopted in our
approach. In section IV, the arbitrary constant amplitudes of the
perturbation expansion are renormalized such as to eliminate the
secular terms. As a result, a set of equations for the
renormalized slowly varying amplitudes is obtained, known as the
renormalization group equations (RGEs). These equations comprise
an infinite system of coupled nonlinear Schr\"{o}dinger equations.
In section V, the latter are analyzed in the simplest case.
Finally, section VI is dedicated to discussion and conclusions.

\renewcommand{\theequation}{\thesection.\arabic{equation}}

\setcounter{equation}{0}

\section{Formulation of the Problem and Basic Equations}

Plasmas and beam-plasma systems considered in the present paper
are assumed to be weakly collisional. Therefore, the dynamics of
plasma species is well described by the hydrodynamic equations
coupled with the equations for the electromagnetic self-fields. We
start with the equations for plasma in an external constant
magnetic field ${\bf B}_0$, which can be written as follows
\begin{equation}
{\frac {\partial n_a} {\partial t}} + \nabla \cdot {\left( n_a
{\bf V}_a \right)} = 0, \label{Continuity}
\end{equation}
\begin{equation}
{\frac {{\rm D}_a {\bf V}_a} {{\rm D} t}} = - {\frac {k_B T_a}
{m_a n_a}} \nabla n_a + {\frac {e q_a} {m_a}} {\left[ {\bf E} +
{\bf V}_a \times {\left( {\bf B}_0 + {\bf B} \right)} \right]},
\label{Mombalance}
\end{equation}
where $n_a$ and ${\bf V}_a$ are the density and the current
velocity of the species $a$. Furthermore, $m_a$, $q_a$ and $T_a$
are the mass, the relative charge and the temperature,
respectively, while $k_B$ is the Boltzmann constant. The
substantional derivative on the left-hand-side of equation
(\ref{Mombalance}) is defined as
\begin{equation}
{\frac {{\rm D}_a} {{\rm D} t}} = {\frac {\partial} {\partial t}}
+ {\bf V}_a \cdot \nabla. \label{Substant}
\end{equation}
The electromagnetic self-fields ${\bf E}$ and ${\bf B}$ can be
obtained in terms of the electromagnetic vector ${\bf A}$ and
scalar $\varphi$ potentials according to the well-known relations
\begin{equation}
{\bf E} = - \nabla \varphi - {\frac {\partial {\bf A}} {\partial
t}}, \qquad {\bf B} = \nabla \times {\bf A}. \label{Elmagfield}
\end{equation}
The latter satisfy the wave equations
\begin{equation}
\Box {\bf A} = - \mu_0 e \sum \limits_a n_a q_a {\bf V}_a, \qquad
\Box \varphi = - {\frac {e} {\epsilon_0}} \sum \limits_a n_a q_a,
\label{Waveequatvec}
\end{equation}
in the Lorentz gauge
\begin{equation}
{\frac {1} {c^2}} {\frac {\partial \varphi} {\partial t}} + \nabla
\cdot {\bf A} = 0. \label{Lorgauge}
\end{equation}
Here $\Box$ denotes the well-known d'Alembert operator. In what
follows, we consider the case of a quasineutral plasma
\begin{equation}
\sum \limits_a n_a q_a = 0, \label{Quasineutr}
\end{equation}
in a constant external magnetic field along the $x$-axis ${\bf
B}_0 = {\left( B_0, 0, 0 \right)}$. Then, equations
(\ref{Continuity})--(\ref{Lorgauge}) possess a stationary solution
\begin{equation}
n_a = n_{a0} = {\rm const}, \quad {\bf V}_a = 0, \quad {\bf A} =
0, \quad \varphi = 0. \label{Statsol}
\end{equation}
The frequency of the wave will be taken as much higher than the
ion-cyclotron frequency. Therefore, we can further neglect the ion
motion and scale the hydrodynamic and field variables as
\begin{equation}
n_e = n_0 + \epsilon N, \quad {\bf V}_e = \epsilon {\bf V}, \quad
{\bf A} \longrightarrow \epsilon {\bf A}, \quad \varphi
\longrightarrow \epsilon \varphi, \label{Scale}
\end{equation}
where $\epsilon$ is a formal small parameter introduced for
convenience, which will be set equal to one at the end of the
calculations. Thus, the basic equations to be used for the
subsequent analysis can be written in the form
\begin{equation}
{\frac {\partial N} {\partial t}} + n_0 \nabla \cdot {\bf V} +
\epsilon \nabla \cdot {\left( N {\bf V} \right)} = 0,
\label{Continuit}
\end{equation}
\begin{equation}
{\frac {\partial {\bf V}} {\partial t}} + \epsilon {\bf V} \cdot
\nabla {\bf V} = - {\frac {k_B T} {m {\left( n_0 + \epsilon N
\right)}}} \nabla N \nonumber
\end{equation}
\begin{equation}
- {\frac {e} {m}} {\left[ {\bf E} + {\bf V} \times {\left( {\bf
B}_0 + \epsilon {\bf B} \right)} \right]}, \label{Mombalanc}
\end{equation}
\begin{equation}
\Box {\bf A} = \mu_0 e {\left( n_0 + \epsilon N \right)} {\bf V},
\qquad {\frac {1} {c^2}} {\frac {\partial \varphi} {\partial t}} +
\nabla \cdot {\bf A} = 0. \label{Waveequat}
\end{equation}
Before we continue with the renormalization group reduction of the
system of equations (\ref{Continuit})--(\ref{Waveequat}) in the
next section, let us assume that the actual dependence of the
quantities $N$, ${\bf V}$, ${\bf A}$ and $\varphi$ on the spatial
variables is represented by the expression
\begin{equation}
{\widehat{\Psi}} = {\widehat{\Psi}} {\left( {\bf x}, {\bf X}; t
\right)}, \qquad {\widehat{\Psi}} = {\left( N, {\bf V}, {\bf A},
\varphi \right)}, \label{Actualdep}
\end{equation}
where ${\bf X} = \epsilon {\bf x}$ is a slow spatial variable.
Thus, the only renormalization parameter left at our disposal is
the time $t$ which will prove extremely convenient and simplify
tedious algebra in the sequel.

\renewcommand{\theequation}{\thesection.\arabic{equation}}

\setcounter{equation}{0}

\section{Renormalization Group Reduction of the Magnetohydrodynamic
Equations}

Following the standard procedure of the renormalization group
method, we represent ${\widehat{\Psi}}$ as a perturbation
expansion
\begin{equation}
{\widehat{\Psi}} = \sum \limits_{n=0}^{\infty} \epsilon^n
{\widehat{\Psi}}_n, \label{Perturbexp}
\end{equation}
in the formal small parameter $\epsilon$. The next step consists
in expanding the system of hydrodynamic and field equations
(\ref{Continuit})-(\ref{Waveequat}) in the small parameter
$\epsilon$, and obtaining their naive perturbation solution order
by order. Note that in all orders the perturbation equations
acquire the general form
\begin{equation}
{\frac {\partial N_n} {\partial t}} + n_0 \nabla \cdot {\bf V}_n =
\alpha_n, \label{Continuitn}
\end{equation}
\begin{equation}
{\frac {\partial {\bf V}_n} {\partial t}} = - {\frac {v_T^2}
{n_0}} \nabla N_n - {\frac {e} {m}} {\bf E}_n - \omega_c {\bf V}_n
\times {\bf e}_x + {\bf W}_n, \label{Mombalancn}
\end{equation}
\begin{equation}
\Box {\bf A}_n = \mu_0 e n_0 {\bf V}_n + {\bf U}_n, \qquad {\frac
{1} {c^2}} {\frac {\partial \varphi_n} {\partial t}} + \nabla
\cdot {\bf A}_n = \beta_n, \label{Waveequatn}
\end{equation}
where $\alpha_n$, $\beta_n$, ${\bf U}_n$ and ${\bf W}_n$ are
quantities, that have been already determined from previous
orders. Here
\begin{equation}
v_T^2 = {\frac {k_B T} {m}}, \qquad \omega_c = {\frac {e B_0} {m}}
\label{Parameters}
\end{equation}
are the thermal velocity of electrons and the electron-cyclotron
frequency, respectively and ${\bf e}_x = {\left( 1, 0, 0 \right)}$
is the unit vector in the $x$-direction. Manipulating in an
obvious manner equations (\ref{Continuitn})--(\ref{Waveequatn}),
it is possible to obtain a single equation for ${\bf A}_n$. The
latter reads as
\begin{equation}
\Box {\frac {\partial^2 {\bf A}_n} {\partial t^2}} - v_T^2 \Box
\nabla {\left( \nabla \cdot {\bf A}_n \right)} + \omega_c \Box
{\frac {\partial {\bf A}_n} {\partial t}} \times {\bf e}_x
\nonumber
\end{equation}
\begin{equation}
- {\frac {\omega_p^2} {c^2}} {\frac {\partial^2 {\bf A}_n}
{\partial t^2}} + \omega_p^2 \nabla {\left( \nabla \cdot {\bf A}_n
\right)} = \mu_0 e n_0 {\frac {\partial {\bf W}_n} {\partial t}} +
{\frac {\partial^2 {\bf U}_n} {\partial t^2}} \nonumber
\end{equation}
\begin{equation}
- \mu_0 e v_T^2 \nabla \alpha_n - v_T^2 \nabla {\left( \nabla
\cdot {\bf U}_n \right)} + \omega_c {\frac {\partial {\bf U}_n}
{\partial t}} \times {\bf e}_x + \omega_p^2 \nabla \beta_n,
\label{Waveeqallord}
\end{equation}
where
\begin{equation}
\omega_p^2 = {\frac {e^2 n_0} {\epsilon_0 m}}, \label{Plasmafreq}
\end{equation}
is the electron plasma frequency. Note that the thermal velocity
$v_T$ as defined by equation (\ref{Parameters}) can be
alternatively expressed according to the expression
\begin{equation}
v_T = \omega_p r_D, \qquad r_D^2 = {\frac {\epsilon_0 k_B T} {e^2
n_0}}, \label{Thermvel}
\end{equation}
where $r_D$ is the electron Debye radius. Equation
(\ref{Waveeqallord}) represents the starting point for the
renormalization group reduction, the final goal of which is to
obtain a description of the relatively slow dynamics leading to
formation of patterns and coherent structures.

Let us proceed order by order. We assume that the dependence on
the fast spatial variables ${\bf x} = {\left( x, y, z \right)}$ is
through the longitudinal (parallel to the external magnetic field
${\bf B}_0$) $x$-coordinate only. The solution to the zero-order
perturbation equations (\ref{Waveeqallord}) can be written as
\begin{equation}
{\bf A}_0 = \sum \limits_{k} {\bf A}_{k}^{(0)} {\cal A}_{k} {\rm
e}^{i \psi_{k}}, \label{Zeroordera}
\end{equation}
where
\begin{equation}
\psi_{k} {\left( x; t \right)} = k x - \omega_{k} t, \label{Phase}
\end{equation}
and ${\cal A}_{k}$ is an infinite set of constant complex
amplitudes, which will be the subject of the renormalization
procedure in the sequel. Here "constant" means that the amplitudes
${\cal A}_{k}$ do not depend on the fast spatial variable $x$ and
on the time $t$, however, it can depend on the slow spatial
variables ${\bf X}$. The summation sign in equation
(\ref{Zeroordera}) and throughout the paper implies summation over
the wave number $k$ in the case where it takes discrete values, or
integration in the continuous case. From the dispersion equation
\begin{equation}
{\cal D} {\left( k; \omega_{k} \right)} = \omega_k^2 {\left[
\omega_k^2 {\left( \Box_k - {\frac {\omega_p^2} {c^2}} \right)}^2
- \omega_c^2 \Box_k^2 \right]} = 0, \label{Disperequat}
\end{equation}
it follows that the wave frequency $\omega_{k}$ can be expressed
in terms of the wave number $k$, where the Fourier-image $\Box_k$
of the d'Alembert operator can be written according to
\begin{equation}
\Box_{k} = {\frac {\omega_{k}^2} {c^2}} - k^2. \label{Dalembert}
\end{equation}
Moreover, it can be verified in a straightforward manner that the
constant vector ${\bf A}_{k}^{(0)}$ can be expressed as
\begin{equation}
{\bf A}_{k}^{(0)} = {\left( 0, 1, -i {\rm sgn} (k) \right)},
\label{Constvect}
\end{equation}
where ${\rm sgn} (k)$ is the well-known sign-function. Details
concerning the derivation of the dispersion law
(\ref{Disperequat}) and equation (\ref{Constvect}) can be found in
the Appendix. Note that equation (\ref{Constvect}) is an
alternative representation of the solvability condition
(\ref{Appsolcond}). It is important to emphasize that
\begin{equation}
\omega_{-k} = - \omega_k, \qquad {\cal A}_{-k} = {\cal
A}_{k}^{\ast}, \label{Importnote}
\end{equation}
where the asterisk denotes complex conjugation. The latter assures
that the vector potential as defined by equation
(\ref{Zeroordera}) is a real quantity. The zero-order current
velocity ${\bf V}_0$ obtained directly from the first equation
(\ref{Waveequatn}) can be written as
\begin{equation}
{\bf V}_0 = \sum \limits_{k} {\bf V}_{k}^{(0)} {\cal A}_{k} {\rm
e}^{i \psi_{k}}, \qquad {\bf V}_{k}^{(0)} = {\frac {\Box_{k}}
{\mu_0 e n_0}} {\bf A}_{k}^{(0)}. \label{Zeroorderv}
\end{equation}
In addition, the zero-order density, scalar potential and magnetic
field are represented by the expressions
\begin{equation}
N_0 \equiv 0, \qquad \varphi_0 \equiv 0, \qquad {\bf B}_0 = \sum
\limits_{k} {\bf B}_{k}^{(0)} {\cal A}_{k} {\rm e}^{i \psi_{k}},
\label{Zeroordern}
\end{equation}
where
\begin{equation}
{\bf B}_{k}^{(0)} = -k {\bf A}_{k}^{(0)} {\rm sgn} (k) = {\left(
0, -k {\rm sgn} (k), ik \right)}. \label{Zeroorderb}
\end{equation}

It has been mentioned that the first-order "source terms" on the
right-hand-side of equation (\ref{Waveeqallord}) can be expressed
via quantities already known from zero order. Thus, we have
\begin{equation}
\alpha_1 = - n_0 {\widehat{\nabla}} \cdot {\bf V}_0, \qquad
\beta_1 = - {\widehat{\nabla}} \cdot {\bf A}_0,
\label{Firstordalp}
\end{equation}
\begin{equation}
{\bf U}_1 = - 2 \nabla \cdot {\widehat{\nabla}} {\bf A}_0, \qquad
{\bf W}_1 = - {\frac {e} {m}} {\bf V}_0 \times {\bf B}_0,
\label{Firstordu}
\end{equation}
where the shorthand notation
\begin{equation}
{\widehat{\nabla}} = {\frac {\partial} {\partial {\bf X}}}
\label{Firstorddef}
\end{equation}
has been introduced. Note that the vector ${\bf W}_1$ representing
the zero-order Lorentz force has the only nonzero component along
the external magnetic field, that is
\begin{equation}
{\bf W}_1 = {\bf e}_x \sum \limits_{k,l} \alpha_{kl} {\cal A}_k
{\cal A}_l {\rm e}^{i {\left( \psi_k + \psi_l \right)}},
\label{Zeroordlf}
\end{equation}
where
\begin{equation}
\alpha_{kl} = - {\frac {i} {2 \mu_0 n_0 m}} {\left( k \Box_l + l
\Box_k \right)} {\left[ 1 - {\rm sgn} (k) {\rm sgn} (l) \right]}.
\label{Firstordalph}
\end{equation}

Equation (\ref{Waveeqallord}) has now two types of solutions. The
first is a secular solution linearly dependent on the time
variable in the first-order approximation. As a rule, the highest
power in the renormalization parameter of the secular terms
contained in the standard perturbation expansion is equal to the
corresponding order in the small perturbation parameter. The
second solution of equation (\ref{Waveeqallord}) arising from the
nonlinear interaction between waves in the first order, is
regular. Omitting tedious but standard algebra, we present here
only the result
\begin{equation}
{\bf A}_{1} = \sum \limits_{k} {\widehat{\bf A}}_{k}^{(1)} {\cal
A}_{k} {\rm e}^{i \psi_{k}} + {\bf e}_x \sum \limits_{k,l}
A_{kl}^{(1)} {\cal A}_{k} {\cal A}_{l} {\rm e}^{i {\left( \psi_k +
\psi_l \right)}} , \label{Firstordvps}
\end{equation}
where
\begin{equation}
{\widehat{\bf A}}_{k}^{(1)} = {\left( {\widehat{A}}_{kx}^{(1)}, t
{\widehat{A}}_{ky}^{(1)}, -i t {\widehat{A}}_{ky}^{(1)} {\rm sgn}
(k) \right)}, \label{Firstordvec}
\end{equation}
Some of the details of the calculations are presented in the
Appendix. In explicit form, the components of the vector operator
${\widehat{\bf A}}_{\bf k}^{(1)}$ and those of the infinite matrix
$A_{kl}^{(1)}$ are given by the expressions
\begin{equation}
{\widehat{A}}_{kx}^{(1)} = - {\frac {i k \beta_k} {\gamma_k
\Box_k}} {\widehat{\nabla}}_k, \qquad {\widehat{\nabla}}_k = {\bf
A}_{k}^{(0)} \cdot {\widehat{\nabla}}, \label{Firstordakx}
\end{equation}
\begin{equation}
{\widehat{A}}_{ky}^{(1)} = - {\frac {{\widehat{F}}_{k}} {2
\omega_k \alpha_k {\rm sgn} (k) + \omega_c \chi_k}},
\label{Firstordaky}
\end{equation}
\begin{equation}
A_{kl}^{(1)} = {\frac {e} {2m v_T^2}} {\frac {\omega_k + \omega_l}
{\Box_{kl} {\cal D}_{kl}}} {\left( k \Box_l + l \Box_k \right)}
{\left[ 1 - {\rm sgn} (k) {\rm sgn} (l) \right]},
\label{Firstordaklx}
\end{equation}
where
\begin{equation}
{\widehat{F}}_{k} = 2k \omega_k {\left[ \omega_k {\rm sgn} (k) +
\omega_c \right]} {\widehat{\nabla}}_{X}, \label{Firstordoper}
\end{equation}
\begin{equation}
\Box_{kl} = {\frac {{\left( \omega_k + \omega_l \right)}^2} {c^2}}
- (k+l)^2, \label{Firstordconst}
\end{equation}
\begin{equation}
{\cal D}_{kl} = {\frac {{\left( \omega_k + \omega_l \right)}^2}
{v_T^2}} - (k+l)^2 - {\frac {1} {r_D^2}}. \label{Firstordconsta}
\end{equation}
In addition, the constants $\alpha_k$, $\beta_k$, $\gamma_k$ and
$\chi_k$ entering the expressions above are given by
\begin{equation}
\alpha_k = \Box_k + {\frac {\omega_k^2 - \omega_p^2} {c^2}},
\qquad \beta_k = \Box_k - {\frac {1} {r_D^2}}, \label{Constantsfo}
\end{equation}
\begin{equation}
\gamma_k = {\frac {\omega_k^2} {v_T^2}} - k^2 - {\frac {1}
{r_D^2}}, \qquad \chi_k = \Box_k + {\frac {2 \omega_k^2} {c^2}}.
\label{Constantsfor}
\end{equation}
Furthermore, the first-order  current velocity can be expressed as
\begin{equation}
{\bf V}_{1} = \sum \limits_{k} {\widehat{\bf V}}_{k}^{(1)} {\cal
A}_{k} {\rm e}^{i \psi_{k}} + {\bf e}_x \sum \limits_{k,l}
V_{kl}^{(1)} {\cal A}_{k} {\cal A}_{l} {\rm e}^{i {\left( \psi_k +
\psi_l \right)}}, \label{Firstordcvel}
\end{equation}
where
\begin{equation}
{\widehat{\bf V}}_{k}^{(1)} = {\left( {\widehat{V}}_{kx}^{(1)},
{\widehat{V}}_{ky}^{(1)}, -i {\widehat{V}}_{ky}^{(1)} {\rm sgn}
(k) \right)}. \label{Firstordvcvel}
\end{equation}
The corresponding operators and matrix coefficients can be written
explicitly according to the expressions
\begin{equation}
{\widehat{V}}_{kx}^{(1)} = {\frac {\Box_k} {\mu_0 e n_0}}
{\widehat{A}}_{kx}^{(1)}, \qquad V_{kl}^{(1)} = {\frac {\Box_{kl}}
{\mu_0 e n_0}} A_{kl}^{(1)}, \label{Currentvelx}
\end{equation}
\begin{equation}
{\widehat{V}}_{ky}^{(1)} = {\frac {1} {\mu_0 e n_0}} {\left[ t
\Box_k {\widehat{A}}_{ky}^{(1)} + 2i {\left( {\frac {\omega_k}
{c^2}} {\widehat{A}}_{ky}^{(1)} + k {\widehat{\nabla}}_X \right)}
\right]}, \label{Currentvely}
\end{equation}
Calculating the first-order density $N_1$ from equation
(\ref{Continuitn}), we obtain
\begin{equation}
N_{1} = \sum \limits_{k} {\widehat{N}}_{k}^{(1)} {\cal A}_{k} {\rm
e}^{i \psi_{k}} + \sum \limits_{k,l} N_{kl}^{(1)} {\cal A}_{k}
{\cal A}_{l} {\rm e}^{i {\left( \psi_k + \psi_l \right)}},
\label{Firstordden}
\end{equation}
\begin{equation}
{\widehat{N}}_{k}^{(1)} = {\frac {\Box_k} {\mu_0 e \omega_k}}
{\left( k {\widehat{A}}_{kx}^{(1)} - i {\widehat{\nabla}}_k
\right)}, \label{Firstorddenc}
\end{equation}
\begin{equation}
N_{kl}^{(1)} = {\frac {k + l} {2 \mu_0 m v_T^2 {\cal D}_{kl}}}
{\left( k \Box_l + l \Box_k \right)} {\left[ 1 - {\rm sgn} (k)
{\rm sgn} (l) \right]}. \label{Firstorddenco}
\end{equation}
Analogously, for the first-order scalar potential $\varphi_1$, we
find
\begin{equation}
\varphi_1 = \sum \limits_{k} {\widehat{\varphi}}_{k}^{(1)} {\cal
A}_{k} {\rm e}^{i \psi_{k}} + \sum \limits_{k,l}
\varphi_{kl}^{(1)} {\cal A}_{k} {\cal A}_{l} {\rm e}^{i {\left(
\psi_k + \psi_l \right)}}, \label{Firstordscp}
\end{equation}
\begin{equation}
{\widehat{\varphi}}_{k}^{(1)} = {\frac {e} {\epsilon_0 \Box_k}}
{\widehat{N}}_{k}^{(1)} = {\frac {c^2} {\omega_k}} {\left( k
{\widehat{A}}_{kx}^{(1)} - i {\widehat{\nabla}}_k \right)},
\label{Firstordscpc}
\end{equation}
\begin{equation}
\varphi_{kl}^{(1)} = {\frac {e c^2 (k+l)} {2 m v_T^2 \Box_{kl}
{\cal D}_{kl}}} {\left( k \Box_l + l \Box_k \right)} {\left[ 1 -
{\rm sgn} (k) {\rm sgn} (l) \right]}. \label{Firstordscpco}
\end{equation}
Finally, the first-order magnetic field is calculated to be
\begin{equation}
{\bf B}_{1} = \sum \limits_{k} {\widehat{\bf B}}_{k}^{(1)} {\cal
A}_{k} {\rm e}^{i \psi_{k}}, \label{Firstordmagf}
\end{equation}
where
\begin{equation}
{\widehat{\bf B}}_{k}^{(1)} = {\left( - i {\rm sgn} (k)
{\widehat{\nabla}}_k, {\widehat{B}}_{ky}^{(1)}, -i
{\widehat{B}}_{ky}^{(1)} {\rm sgn} (k) \right)},
\label{Firstordmagfi}
\end{equation}
\begin{equation}
{\widehat{B}}_{ky}^{(1)} = - {\rm sgn} (k) {\left( t k
{\widehat{A}}_{ky}^{(1)} - i {\widehat{\nabla}}_X \right)}.
\label{Firstordmafi}
\end{equation}

A couple of interesting features of the zero and first-order
perturbation solution are noteworthy to be commented at this
point. First of all, the zero-order density $N_0$ vanishes which
means that no density waves are induced by the whistler
eigenmodes. The second terms in the expressions for the
first-order density $N_1$ and current velocity ${\bf V}_1$ [see
equations (\ref{Firstordcvel}) and (\ref{Firstordden})] imply
contribution from nonlinear interaction between waves according to
the nonlinear Lorentz force. It will be shown in the remainder
that these terms give rise to nonlinear terms in the
renormalization group equation and describe solitary wave
behaviour of the whistler mode.

\renewcommand{\theequation}{\thesection.\arabic{equation}}

\setcounter{equation}{0}

\section{The Renormalization Group Equation}

Passing over to the final stage of our renormalization group
procedure, we note that in second order the quantities ${\bf U}_2$
and ${\bf W}_2$ entering the right-hand-side of equation
(\ref{Waveeqallord}) can be written as
\begin{equation}
{\bf U}_2 = - 2 \nabla \cdot {\widehat{\nabla}} {\bf A}_1 -
{\widehat{\nabla}}^2 {\bf A}_0 + \mu_0 e N_1 {\bf V}_0,
\label{Secondordu}
\end{equation}
\begin{equation}
{\bf W}_2 = {\frac {e} {m}} {\widehat{\nabla}} \varphi_1 - {\frac
{v_T^2} {n_0}} {\widehat{\nabla}} N_1 - {\bf V}_1 \cdot \nabla
{\bf V}_0 - {\frac {e} {m}} {\bf V}_1 \times {\bf B}_0,
\label{Secondordw}
\end{equation}
Since we are interested only in the secular terms in second order,
appearing in the expressions for the $y$ and $z$ components of the
electromagnetic vector potential ${\bf A}_2$, contributions in the
source vectors ${\bf U}_2$ and ${\bf W}_2$ leading to such terms
are sufficient for completing the renormalization group procedure.
Thus, we can write
\begin{equation}
{\bf A}_2 = \sum \limits_{k} {\left( t {\widehat{\bf A}}_k^{(2)} +
t^2 {\widehat{\bf C}}_k \right)} {\cal A}_{k} {\rm e}^{i \psi_{k}}
\nonumber
\end{equation}
\begin{equation}
+ t \sum \limits_{k} {\widehat{\bf D}}_k^{(2)} {\cal A}_{k} {\rm
e}^{i \psi_{k}} + t \sum \limits_{k,l} {\mathbf \Gamma}_{kl}
{\left| {\cal A}_{l} \right|}^2 {\cal A}_{k} {\rm e}^{i \psi_k}.
\label{Secondordvps}
\end{equation}
An important remark is in order at this point. From the
solvability condition (\ref{Appsolcond}) it follows that the
complex amplitude ${\cal A}_k$ must satisfy the complex Poisson
equation
\begin{equation}
{\widehat{\nabla}}_k^2 {\cal A}_k = 0. \label{Secondordscon}
\end{equation}
The latter imposes additional restrictions on the dependence of
the wave amplitudes ${\cal A}_k$ on the slow transverse
independent variables $Y$ and $Z$. Straightforward calculations
yield (see the Appendix for details)
\begin{equation}
{\widehat{A}}_{ky}^{(2)} = - {\frac {i {\rm sgn} (k)} {2 \omega_k
\alpha_k {\rm sgn} (k) + \omega_c \chi_k}} {\left( \beta_k^{(2)}
{\widehat{A}}_{ky}^{(1) {\bf 2}} - {\widehat{G}}_k \right)},
\label{Secondordaky}
\end{equation}
\begin{equation}
{\widehat{D}}_{ky}^{(2)} = {\frac {i v_T^2 \beta_k {\rm sgn} (k)}
{2 \omega_k \alpha_k {\rm sgn} (k) + \omega_c \chi_k}} {\left( 1 +
{\frac {k^2 \beta_k} {\gamma_k \Box_k}} \right)}
{\widehat{\nabla}}_Y {\widehat{\nabla}}_k, \label{Secondorddky}
\end{equation}
\begin{equation}
{\widehat{C}}_{ky} = {\frac {1} {2}} {\widehat{A}}_{ky}^{(1) {\bf
2}}, \label{Secondordcky}
\end{equation}
where
\begin{equation}
\beta_k^{(2)} = \alpha_k + {\frac {4 \omega_k^2} {c^2}} + {\frac
{3 \omega_c \omega_k} {c^2}} {\rm sgn} (k), \label{Secondordcon}
\end{equation}
\begin{equation}
{\widehat{G}}_k = \omega_k {\rm sgn} (k) {\left[ \omega_k {\rm
sgn} (k) + \omega_c \right]} {\widehat{\nabla}}^2.
\label{Secondordoper}
\end{equation}
The matrix coefficient $\Gamma_{kly}$ determining the nonlinear
contribution represented by the second term in equation
(\ref{Secondordvps}) reads explicitly as
\begin{equation}
\Gamma_{kly} = - {\frac {1 - {\rm sgn} (k) {\rm sgn} (l)} {\mu_0
n_0 m v_T^2 \omega_l {\cal D}_{kl}}} {\frac {i \omega_k \Box_l
{\left( k \Box_l + l \Box_k \right)} {\rm sgn} (k)} {2 \omega_k
\alpha_k {\rm sgn} (k) + \omega_c \chi_k}} \nonumber
\end{equation}
\begin{equation}
\times {\left[ \omega_c {\left( l \omega_k - k \omega_l \right)}
{\rm sgn} (l) + (k+l) \omega_k \omega_l \right]}.
\label{Secondordgam}
\end{equation}
Following the standard procedure \cite{Tzenov} of the RG method,
we finally obtain the desired RG equation
\begin{equation}
{\frac {\partial {\widetilde{\cal A}}_k} {\partial t}} - \epsilon
{\widehat{A}}_{ky}^{(1)} {\widetilde{\cal A}}_k \nonumber
\end{equation}
\begin{equation}
= \epsilon^2 {\left( {\widehat{A}}_{ky}^{(2)} +
{\widehat{D}}_{ky}^{(2)} \right)} {\widetilde{\cal A}}_k +
\epsilon^2 \sum \limits_l \Gamma_{kly} {\left| {\widetilde{\cal
A}}_l \right|}^2 {\widetilde{\cal A}}_k, \label{Rgroupeq}
\end{equation}
where now ${\widetilde{\cal A}}_k$ is the renormalized complex
amplitude \cite{Tzenov}. Thus, the renormalized solution for the
electromagnetic vector potential acquires the form
\begin{equation}
{\bf A} = \sum \limits_{k} {\bf A}_{k}^{(0)} {\widetilde{\cal
A}}_{k} {\rm e}^{i \psi_{k}}. \label{Renormsolut}
\end{equation}
Analogously, for the electric and magnetic field of the whistler
wave, one can obtain in a straightforward manner the following
expressions
\begin{equation}
{\bf B} = \sum \limits_{k} {\bf B}_{k}^{(0)} {\widetilde{\cal
A}}_{k} {\rm e}^{i \psi_{k}}, \qquad {\bf E} = i \sum \limits_{k}
\omega_k {\bf A}_{k}^{(0)} {\widetilde{\cal A}}_{k} {\rm e}^{i
\psi_{k}}. \label{Renormsolb}
\end{equation}

It is important to mention that the plasma density remains
unchanged ($N = 0$) contrary to the case of electrostatic waves,
where the evolution of the induced electrostatic waves follows the
evolution of the density waves.

\renewcommand{\theequation}{\thesection.\arabic{equation}}

\setcounter{equation}{0}

\section{\label{Essent}System of Coupled Nonlinear Schr\"{o}dinger
Equations}

The simplest case of the validity of the solvability condition
(\ref{Secondordscon}) consists in the assumption that the slow
wave amplitudes ${\cal A}_k$ do not depend on the transverse
coordinates. Setting $\epsilon = 1$ in equation (\ref{Rgroupeq}),
we obtain the following system of coupled nonlinear
Schr\"{o}dinger equations
\begin{equation}
i {\rm sgn} (k) {\frac {\partial {\cal A}_k} {\partial t}} + i
\nu_k {\rm sgn} (k) {\frac {\partial {\cal A}_k} {\partial x}} =
\lambda_k {\frac {\partial^2 {\cal A}_k} {\partial x^2}} + \sum
\limits_l \mu_{kl} {\left| {\cal A}_l \right|}^2 {\cal A}_k,
\label{Couplednse}
\end{equation}
where for simplicity the tilde-sign over the renormalized
amplitude has been dropped. Moreover, the coefficients $\nu_k$,
$\lambda_k$ and $\mu_{kl}$ are given by the expressions
\begin{equation}
\nu_k = {\frac {2k \omega_k {\left[ \omega_k {\rm sgn} (k) +
\omega_c \right]}} {2 \omega_k \alpha_k {\rm sgn} (k) + \omega_c
\chi_k}}, \label{Coefficientnu}
\end{equation}
\begin{equation}
\lambda_k = {\frac {\omega_k {\left[ \omega_k {\rm sgn} (k) +
\omega_c \right]}} {2 \omega_k \alpha_k {\rm sgn} (k) + \omega_c
\chi_k}} \nonumber
\end{equation}
\begin{equation}
\times {\left\{ {\frac {4k^2 \omega_k \beta_k^{(2)} {\left[
\omega_k {\rm sgn} (k) + \omega_c \right]}} {{\left[ 2 \omega_k
\alpha_k {\rm sgn} (k) + \omega_c \chi_k \right]}^2}} - {\rm sgn}
(k) \right\}}, \label{Coefficientla}
\end{equation}
\begin{equation}
\mu_{kl} = {\frac {1 - {\rm sgn} (k) {\rm sgn} (l)} {\mu_0 n_0 m
v_T^2 \omega_l {\cal D}_{kl}}} {\frac {\omega_k \Box_l {\left( k
\Box_l + l \Box_k \right)}} {2 \omega_k \alpha_k {\rm sgn} (k) +
\omega_c \chi_k}} \nonumber
\end{equation}
\begin{equation}
\times {\left[ \omega_c {\left( l \omega_k - k \omega_l \right)}
{\rm sgn} (l) + (k+l) \omega_k \omega_l \right]}.
\label{Coefficientmu}
\end{equation}
Interestingly enough, the infinite matrix of coupling coefficients
$\mu_{kl}$ represents a sort of selection rules. Clearly,
\begin{equation}
\mu_{kk} = 0, \qquad \mu_{k, -k} = 0, \label{Selectrule1}
\end{equation}
and
\begin{equation}
\mu_{kl} = 0, \qquad {\rm for} \quad {\rm sgn} (k) {\rm sgn} (l) =
1. \label{Selectrule2}
\end{equation}
This means that a generic mode with a wave number $k$ cannot
couple with itself, neither can it couple with another mode with a
wave number of the same sign. Note that this feature is a
consequence of the vector character of the nonlinear coupling
between modes and is due to the nonlinear Lorentz force.
Therefore, for a given mode $k$ the simplest nontrivial reduction
of the infinite system of coupled nonlinear Schr\"{o}dinger
equations consists of minimum two coupled equations.

Without loss of generality, we can assume in what follows that the
sign of an arbitrary mode $k$ under consideration is positive ($k
> 0$). Suppose that for a particular whistler mode with a
positive wave number $k$ there exist a mode with wave number $-l$
for which the coupling coefficient $\mu_{k, -l}$ is maximum.
Neglecting all other modes but the modes $k$ and $-l$, we can
write
\begin{equation}
i {\frac {\partial {\cal A}_k} {\partial t}} + i \nu_k {\frac
{\partial {\cal A}_k} {\partial x}} = \lambda_k {\frac {\partial^2
{\cal A}_k} {\partial x^2}} + \mu_1 {\left| {\cal A}_l \right|}^2
{\cal A}_k, \label{Couplednsek}
\end{equation}
\begin{equation}
i {\frac {\partial {\cal A}_l} {\partial t}} + i \nu_l {\frac
{\partial {\cal A}_l} {\partial x}} = \lambda_l {\frac {\partial^2
{\cal A}_l} {\partial x^2}} + \mu_2 {\left| {\cal A}_k \right|}^2
{\cal A}_l, \label{Couplednsel}
\end{equation}
where
\begin{equation}
\mu_1 = {\frac {2} {\mu_0 n_0 m v_T^2 \omega_l {\cal D}_{k, -l}}}
{\frac {\omega_k \Box_l {\left( k \Box_l - l \Box_k \right)}} {2
\omega_k \alpha_k + \omega_c \chi_k}} \nonumber
\end{equation}
\begin{equation}
\times {\left[ \omega_c {\left( k \omega_l - l \omega_k \right)} +
(k-l) \omega_k \omega_l \right]}. \label{Coefficientmu1}
\end{equation}
\begin{equation}
\mu_2 = {\frac {2} {\mu_0 n_0 m v_T^2 \omega_k {\cal D}_{k, -l}}}
{\frac {\omega_l \Box_k {\left( k \Box_l - l \Box_k \right)}} {2
\omega_l \alpha_l + \omega_c \chi_l}} \nonumber
\end{equation}
\begin{equation}
\times {\left[ \omega_c {\left( k \omega_l - l \omega_k \right)} +
(k-l) \omega_k \omega_l \right]}. \label{Coefficientmu2}
\end{equation}

The system of coupled nonlinear Schr\"{o}dinger equations
(\ref{Couplednsek}) and (\ref{Couplednsel}) is non integrable in
general \cite{Manakov}. It represents an important starting point
for further investigations on the nonlinear dynamics and evolution
of whistler waves in magnetized plasmas.

\renewcommand{\theequation}{\thesection.\arabic{equation}}

\setcounter{equation}{0}

\section{Discussion and conclusions}

We studied the nonlinear dynamics of whistler waves in magnetized
plasmas. Since plasmas and beam-plasma systems considered here are
assumed to be weakly collisional, the point of reference for the
analysis performed in the present paper is the system of
hydrodynamic and field equations. We apply the renormalization
group method to obtain dynamical equations for the slowly varying
amplitudes of whistler waves. As a result of the investigation
performed, it has been shown that the amplitudes of eigenmodes
satisfy an infinite system of coupled nonlinear Schr\"{o}dinger
equations.  In this sense, the whistler eigenmodes form a sort of
a gas of interacting quasiparticles, while the slowly varying
amplitudes can be considered as dynamical variables heralding the
relevant information about the system.

An important feature of our description is that whistler waves do
not perturb the initial uniform density of plasma electrons. The
plasma response to the induced whistler waves consists in velocity
redistribution which follows exactly the behaviour of the
whistlers. Another interesting peculiarity are the selection rules
governing the nonlinear mode coupling. According to these rules
modes with the same sign do not couple, which is a direct
consequence of the vector character of the interaction. Careful
inspection shows that the initial source of the nonlinear
interaction between waves in the whistler mode is the zero-order
Lorentz force [see equation (\ref{Zeroordlf})]. Since the quantity
${\bf W}_1$ is proportional to ${\bf A}_k^{(0)} \times {\bf
A}_l^{(0)}$, the above mentioned selection rules follow directly,
provided the only case in which the cross product does not vanish
is the case, where modes $k$ and $l$ have different sign.

We believe that the results obtained in the present paper might
have a wide class of possible applications ranging from laboratory
experiments to observations of a variety of effects relevant to
space plasmas.

\begin{acknowledgments}
It is a pleasure to thank B. Baizakov for many interesting and
useful discussions concerning the subject of the present paper.
\end{acknowledgments}

\appendix

\renewcommand{\theequation}{\thesection.\arabic{equation}}

\setcounter{equation}{0}

\section{\label{app:subsec} Details Concerning the Derivation of
the Perturbation Expansion}

Under the assumption that whistler waves propagate parallel to the
external magnetic field ${\bf B}_0$ (dependence on the
longitudinal $x$-coordinate and the time $t$ only), the equation
(\ref{Waveeqallord}) for the longitudinal component $A_{nx}$
decouple from the equations for the other two components. In zero
order it has an obvious solution $A_{0x} = 0$. The perturbation
equations for the transverse components of the electromagnetic
vector potential can be written as
\begin{equation}
\Box {\frac {\partial^2 A_{0y}} {\partial t^2}} + \omega_c \Box
{\frac {\partial A_{0z}} {\partial t}} - {\frac {\omega_p^2}
{c^2}} {\frac {\partial^2 A_{0y}} {\partial t^2}} = 0,
\label{Appzeroy}
\end{equation}
\begin{equation}
\Box {\frac {\partial^2 A_{0z}} {\partial t^2}} - \omega_c \Box
{\frac {\partial A_{0y}} {\partial t}} - {\frac {\omega_p^2}
{c^2}} {\frac {\partial^2 A_{0z}} {\partial t^2}} = 0.
\label{Appzeroz}
\end{equation}
To solve the above system of equations, we use the ansatz
(\ref{Zeroordera}). Then, the dispersion equation
(\ref{Disperequat}) can be obtained in a straightforward manner as
a condition for vanishing of the determinant of the linear system
consisting of the components $A_{ky}^{(0)}$ and $A_{kz}^{(0)}$.
Moreover, from the dispersion equation it follows that
$A_{kz}^{(0)}$ is proportional to $A_{ky}^{(0)}$, that is
\begin{equation}
A_{kz}^{(0)} = -i {\rm sgn} (k) A_{ky}^{(0)}. \label{Appsolcond}
\end{equation}
Equation (\ref{Appsolcond}) should be regarded as a solvability
condition and must be satisfied order by order.

The first order is characterized by the presence of resonant terms
(proportional to ${\rm e}^{i \psi_k}$) on the right-hand-side of
equation (\ref{Waveeqallord}). These yield a secular solution to
the perturbation equations linear (in first order) with respect to
the time $t$. Taking into account the resonant terms alone, it can
be verified that the operators ${\widehat{A}}_{ky}^{(1)}$ and
${\widehat{A}}_{kz}^{(1)}$ satisfy the following system of linear
equations
\begin{equation}
2i \omega_k \alpha_k {\widehat{A}}_{ky}^{(1)} - \omega_c \chi_k
{\widehat{A}}_{kz}^{(1)} = - i {\rm sgn} (k) {\widehat{F}}_k,
\label{Appfolinsysy}
\end{equation}
\begin{equation}
2i \omega_k \alpha_k {\widehat{A}}_{kz}^{(1)} + \omega_c \chi_k
{\widehat{A}}_{ky}^{(1)} = - {\widehat{F}}_k, \label{Appfolinsysz}
\end{equation}
The solvability condition (\ref{Appsolcond}) which also holds in
first order yields immediately equations (\ref{Firstordvec}) and
(\ref{Firstordaky}). Non resonant terms can be handled in a
straightforward manner, yielding the second term on the
right-hand-side of equation (\ref{Firstordvps}).

In second order the right-hand-side of equation
(\ref{Waveeqallord}) contains resonant terms proportional to ${\rm
e}^{i \psi_k}$, as well as terms proportional to $t {\rm e}^{i
\psi_k}$. Contributions to the second-order solution of the first
type can be handled in a way similar to that already discussed in
first order. To deal with the second type of resonant
contributions, we write the second-order perturbation equations as
\begin{equation}
\Box {\frac {\partial^2 A_{2y}} {\partial t^2}} + \omega_c \Box
{\frac {\partial A_{2z}} {\partial t}} - {\frac {\omega_p^2}
{c^2}} {\frac {\partial^2 A_{2y}} {\partial t^2}} = it {\rm sgn}
(k) H_k {\cal A}_k {\rm e}^{i \psi_k}, \label{Appsecondy}
\end{equation}
\begin{equation}
\Box {\frac {\partial^2 A_{2z}} {\partial t^2}} - \omega_c \Box
{\frac {\partial A_{2y}} {\partial t}} - {\frac {\omega_p^2}
{c^2}} {\frac {\partial^2 A_{2z}} {\partial t^2}} = t H_k {\cal
A}_k {\rm e}^{i \psi_k}. \label{Appsecondz}
\end{equation}
It is straightforward to verify that the solution to equations
(\ref{Appsecondy}) and (\ref{Appsecondz}) is of the form
\begin{equation}
A_{2y,z} = {\left( t^2 C_{ky,z} + t L_{ky,z} \right)} {\cal A}_k
{\rm e}^{i \psi_k}, \label{Appsecondgs}
\end{equation}
where the coefficients $C_{ky,z}$ and $L_{ky,z}$ can be written
as
\begin{equation}
C_{ky} = - {\frac {H_k} {2 {\left( 2 \omega_k \alpha_k {\rm sgn}
(k) + \omega_c \chi_k \right)}}}, \label{Appsecondcy}
\end{equation}
\begin{equation}
L_{ky} = -i {\frac {2 \beta_k^{(2)} C_{ky} {\rm sgn} (k)} {2
\omega_k \alpha_k {\rm sgn} (k) + \omega_c \chi_k}},
\label{Appsecondly}
\end{equation}
\begin{equation}
C_{kz} = -i C_{ky} {\rm sgn} (k), \qquad L_{kz} = -i L_{ky} {\rm
sgn} (k). \label{Appsecondclyz}
\end{equation}
Substituting the appropriate form of $H_k$ and collecting similar
terms proportional to $t$ and $t^2$, we readily obtain equations
(\ref{Secondordaky})--(\ref{Secondordcky}).

\end{document}